# Mössbauer, nuclear inelastic scattering and density functional studies on the second metastable state of Na$_2$[Fe(CN)$_5$NO]·2H$_2$O

V. Rusanov,[1] H. Paulsen,[2] L. H. Böttger,[2] H. Winkler,[2] J. A. Wolny,[2] N. Koop,[3]
Th. Dorn,[4] C. Janiak,[4] and A. X. Trautwein[2]

[1]*Department of Atomic Physics, Faculty of Physics, University of Sofia, 1126 Sofia, Bulgaria*
[2]*Institut für Physik, Universität zu Lübeck, D-23538 Lübeck, Germany*
[3]*Institut für Biomedizinische Optik, Universität zu Lübeck, D-23538 Lübeck, Germany*
[4]*Institut für Anorganische und Analytische Chemie, Universität Freiburg, D-79104 Freiburg, Germany*

The structure of the light-induced metastable state SII of Na$_2$[Fe(CN)$_5$NO]·2H$_2$O was investigated by transmission Mössbauer spectroscopy (TMS) in the temperature range between 85 and 135 K, nuclear inelastic scattering (NIS) at 98 K using synchrotron radiation and density functional theory (DFT) calculations. The DFT and TMS results strongly support the view that the NO group in SII takes a side-on molecular orientation and, further, is dynamically displaced from one eclipsed, via a staggered, to a second eclipsed orientation. The population conditions for generating SII are optimal for measurements by TMS, yet they are modest for accumulating NIS spectra. Optimization of population conditions for NIS measurements is discussed and new NIS experiments on SII are proposed.

# I. INTRODUCTION

In sodium nitroprusside, $Na_2[Fe(CN)_5NO]\cdot 2H_2O$ (SNP), changes of the optical dispersion occur after illumination with light at 77 K. Hauser and coworkers [1] discovered in 1977 by means of transmission Mössbauer spectroscopy (TMS), that these changes are due to the population of long-lived metastable states of the nitroprusside anion, $[Fe(CN)_5NO]^{2-}$ (NP). A maximum of about 45 % of the anions can be transferred from the ground state (GS) into the first metastable state, SI, by illumination with blue light of 457.9 nm from an $Ar^+$-laser and only a few percent into the second state, SII. The latter state can, however, be populated up to about 25 % when a crystal with SI already populated at maximum is illuminated a second time with infrared light from an Nd-YAG-laser with wavelength 1064 nm. Both states can be completely and reversibly transferred to the ground state by illumination with light in the red spectral range, e.g. of 632.8 nm from a He-Ne-laser, or by (thermally) annealing the sample. The decay temperatures for SII and SI in SNP are approximately 150 K and 200 K, respectively. A summary of the Mössbauer parameters for GS, SI and SII are described elsewhere [2]. The photochromatic effects are very strong: in GS the crystal is red, in SI it is light brown, and in SII it is black. Apart from the low critical temperatures this kind of photochromacity has been proposed as a new method for optical storage of information with an extremely high density [3]. The existence of light-induced long-lived metastable states is typical for the nitrosyl complexes of all transition metals. In two ruthenium complexes the decay temperature for SI lies even at room temperature [4], which makes this complex especially suitable for practical applications.

## A. Structure determination of the metastable states SI and SII

The study of nitroprussides was accompanied by an ongoing controversy concerning the molecular conformation of both metastable states, SI and SII. Neutron diffraction studies have indicated only small structural changes between the exited states SI [5-7], SII [8] and the ground state GS, as for example in the Fe–N bond length (~ 0.05 Å) and in the C–Fe–N angle (~ 3.5°). Such small changes could hardly account for the long lifetimes of these metastable states [9]. In contrast, the X-ray diffraction studies of Coppens and coworkers [10] reveal that in the SI state the NO group is inverted and takes the so-called isonitrosyl structure N−C−Fe−O−N, and in the SII state the NO group takes a side-on molecular conformation.

Electronic structure calculations, by means of density functional theory (DFT), have been performed with the aim to shed light on the geometrical and electronic structure of SI and SII [11-13]. These studies also support the isonitrosyl structure for the SI state and the side-on conformation of the NO group for the SII state. The isonitrosyl structure of SI was also supported by IR and Raman studies using isotope-substituted ($^{15}N$, $^{18}O$, $^{54}Fe$) nitroprussides [14, 15].

Mössbauer measurements have yielded significant line broadening of the SII-quadrupole doublet at about 80 K [16, 17]. This result was taken as indication that the side-on orientation of the NO-group is dynamically disordered.

### B. Nuclear inelastic scattering of GS and SI

Recently we have made use of nuclear inelastic scattering (NIS) of synchrotron radiation to study molecular vibrations of GS and of SI in guanidinium nitroprusside single crystals [18, 19]. NIS is a powerful tool for the investigation of the dynamics of Mössbauer nuclei in molecular crystals. Information about only those normal modes of the vibrational spectrum that are connected with the iron center is achieved [20-22]. DFT simulations of NIS spectra on the basis of the isonitrosyl structure of SI reveal a significant downshift of all molecular vibrational modes as compared with GS, in full agreement with experiment. Therfore, it was motivating to perform comparable NIS studies on the SII state.

### C. Reinvestigation of SI and SII by neutron diffraction

In the light of X-ray, TMS, NIS, DFT, IR, and Raman studies which are supporting the isonitrosyl structure of SI and the side-on conformation of the NO group of SII, neutron diffraction studies of SII [23] and SI [24] were repeated. Special care was taken to determine the population of SI and SII by low-temperature absorption spectroscopy. These neutron diffraction studies are now in line with the above mentioned structural results for SI and SII. The unusually large displacement parameter of the nitrogen atom of SII was attributed to a tunneling process between the eclipsed and the staggered side-on conformation of the NO group.

A dynamic structural disorder of the side-one conformation of NO in SII is expected to display new characteristic iron-ligand vibrations. We therefore have performed Mössbauer (in the temperature range 85 - 135 K), NIS (using synchrotron radiation) and DFT studies on SII in SNP.

## II. EXPERIMENTAL AND COMPUTATIONAL DETAILS

### A. Samples, population conditions and transmission Mössbauer spectroscopy

For the two sets of measurements, TMS and NIS, two different samples were mounted in a small portable, home-made cryostat, allowing sample transportation between three different institutions where optical illuminations (Institute of Biomedical Optics, Lübeck), Mössbauer (Institute of Physics, Lübeck) and NIS (ESRF, Grenoble) investigations were performed. For the TMS measurements a small **b**-cut SNP single crystal (**b**-crystal direction is parallel to the propagation direction of the Mössbauer radiation) with a diameter of 5 mm, crystal thickness 490 μm and natural isotope abundance of iron has been used. For the NIS measurements a small 2x2x6 mm SNP single crystal, enriched to 95 % with the Mössbauer isotope $^{57}$Fe, has been mounted on the cold finger of the cryostat so that the **b**-crystal direction again was parallel to the propagation direction of the synchrotron beam. A detailed description of the sample preparation (synthesis and crystal growth) is given elsewhere [25].

Both crystals were illuminated in the same way. The first illumination (to populate SI from GS) was performed with blue light, 457.9 nm, from an Ar$^+$-laser, wave vector **k**∣**b** and plane of polarization **E**∣**c**-axis, equally from both sides (one day for each side), up to an integrated light exposure of about 35000 J/cm$^2$, using an average power of 200 mW/cm$^2$. The second illumination (to populate SII from SI) was performed with infrared light, 1064 nm, from a Nd-YAG-laser, wave vector **k│b,** plane of polarization **E│c**-axis and an average power of 200 mW/cm$^2$, again equally from both sides, to an average exposure of about 200 J/cm$^2$. TMS measurements [16] have shown that at these conditions a population of SII up to 25 % could be expected.

The Mössbauer (TMS) spectra were accumulated in transmission geometry by using a standard spectrometer working in constant acceleration mode. All isomer shifts are related to the α-Fe standard at room temperature.

### B. Nuclear inelastic scattering of SII

NIS spectra were obtained at the Nuclear Resonance Beamline ID18 [26, 27] of the European Synchrotron Radiation Facility (ESRF) in Grenoble, France. Prior to the experiment in Grenoble, the SNP crystal was populated with SII in a transportable cryostat and kept cold during transportation from Lübeck to Grenoble.

### C. Density functional calculations

Electronic structure calculations were performed applying density functional theory (DFT) in order to identify local minima on the energy hypersurface that correspond to the groundstate and to the metastable states SI, SIIa (eclipsed), SIIb (staggered) and to the transition state SIIts connecting SIIa and SIIb. The hybrid functional B3LYP [28] has been used together with Ahlrich's triple zeta valence basis set with polarization functions [29] (TZVP) as implemented in the program package Gaussian03 [30]. All calculations have been performed for a free nitroprusside anion [Fe(CN)$_5$(NO)]$^{2-}$ *in vacuo*. Full geometry optimizations have been carried out for different conformations of the nitroprusside anion and subsequently the normal modes of molecular vibrations within the harmonic approximation, and the electric field gradient (EFG) at the iron nucleus have been calculated. Tight convergence criteria and an ultrafine integration grid have been used for the optimizations and for the frequency calculations.

## III. RESULTS

### A. DFT results for the eclipsed-staggered dynamic model for SII

DFT calculations predict for the side-on conformation proposed for SII a new line in the NIS spectrum with energy of 67 meV, suitably positioned where no vibrational modes of GS exist. This mode can be described as a combination of Fe-N-O bending and Fe-$C_{ax}$ bond stretching with the iron center mainly moving along the molecular symmetry axis (*i. e.* the axis perpendicular to the equatorial CN groups).

Our DFT studies yield for SII two local minima on the energy hypersurface for the nitroprusside anion with a side-on bonded NO group. One of the two minima corresponds to an eclipsed conformation of the NO group (SIIa), and the second to a staggered conformation (SIIb). These two conformations are close in energy, with E(SIIb) being slightly higher than E(SIIa): E(SIIa) − E(SIIb) = −124 K. This energy difference is so small that, within the accuracy of the DFT method and the free molecule approximation, it is not possible to decide which of the two calculated minima actually would be populated at very low temperature (e.g. at 4 K).

Since the two conformations, staggered and eclipsed, are so close in energy, the question arises, whether the NO group can switch between these two orientations or even freely rotate about the symmetry axis of the anion, as has been suggested by Delley *et al.* [11] and by Blaha *et al.* [12]. Answering this question requires to estimate the energy barrier on the pathway from SIIa to SIIb. We, therefore, have calculated a transition state SIIts using the synchronous transit-guided quasi-Newton method [31] implemented in the Gaussian03 program package. This transition state is defined by one negative eigenvalue of the Hessian matrix. The eigenvector that corresponds to the negative eigenvalue gives the direction of the pathway from SIIa via SIIts to SIIb and can in good approximation be described as hindered rotation of the NO group. A superposition of the calculated geometries of the two metastable states SIIa and SIIb and of the transition state SIIts is depicted in Fig. 1. This superposition reveals relatively small displacements of the oxygen atom and larger displacements of the nitrogen atom of the NO group. This finding is consistent with the suggestion of Schaniel *et al.* [23] who proposed that tunneling processes between SIIa and SIIb might be responsible for the large displacement parameter of the nitrogen atom (300 pm$^2$) and for the comparatively small value of the oxygen atom (30 pm$^2$), which have been determined by neutron diffraction. The major difference between these three structures (SIIa, SIIts, and SIIb) is the orientation of the NO group, while the calculated iron-ligand bond distances are almost the same for all three states: the Fe–N and the Fe–C$_{ax}$ bond lengths vary by less than 1 pm. Although changes between 1 and 3 pm are calculated for individual equatorial Fe–C bond distances the changes of the averaged Fe–C$_{eq}$ bond lengths are less than 0.4 pm. The total electronic energy of SIIts is 190 K higher than that of SIIa and 66 K higher than that of SIIb.

Regardless of the smallness of the electronic energy difference the contributions of molecular vibrations to the total energy difference and to the free energy difference can not be neglected. For this reason the calculated frequencies have been used to determine the vibrational contributions to the energy and to the entropy at 77 K. It turns out that the vibrational energy of the eclipsed conformation (SIIa) is slightly higher than that of the staggered conformation (SIIb). As overall result SIIb is 13 K below SIIa. If, additionally, the free energy is considered, SIIb is about 200 K below SIIa. At the current level of accuracy, electronic structure calculations do not prove but, at least, support a dynamic displacement of the NO group over the two orientations corresponding to SIIa and SIIb.

For the two local minima we have also calculated the traceless electric field gradient (EFG) tensors $V$ at the iron nucleus. The coordinate system has been chosen such that the equatorial CN ligands coincide with the x and y axes. The NO group is located in the xz plane (eclipsed conformation, SIIa) or in the plane bisecting the xz and the yz planes (staggered conformation, SIIb), respectively. The calculated EFG tensor for SIIa,

$$V(\text{SIIa}) = \begin{vmatrix} -0.2 & 0.0 & -0.2 \\ 0.0 & +1.6 & 0.0 \\ -0.2 & 0.0 & -1.4 \end{vmatrix} \text{ a. u.,}$$

has the eigenvalues –1.44 a.u., –0.16 a.u., and +1.61 a.u. yielding as asymmetry parameter the value $\eta = 0.8$. Taking 0.16 barn for the quadrupole moment of the $^{57}$Fe nucleus we obtain a quadrupole splitting $\Delta E_Q$ of –2.84 mm/s. While the magnitude of the calculated $\Delta E_Q$ is in agreement with the experimental value [16] of about +2.86 mm/s, the calculated negative sign is not in accordance with the experimentally determined positive sign [16] and, further, the calculated $\eta$ is significantly larger than the experimentally observed value of about 0.17(5) [16]. Large asymmetry parameters have also been calculated by Delley *et al.* [11] for a free NP anion ($\eta$=0.93) and for the solid with periodic boundary conditions by Delley *et al.*[11] and by Blaha *et al.* [12] (0.44 and 0.51, respectively). The calculated EFG tensor for SIIb,

$$V(\text{SIIb}) = \begin{vmatrix} +0.8 & -0.5 & 0.0 \\ -0.5 & +0.8 & 0.0 \\ 0.0 & 0.0 & -1.6 \end{vmatrix} \text{ a. u.,}$$

has the eigenvalues –1.63 a.u., +0.30 a.u., and +1.28 a.u., which correspond to a quadrupole splitting $\Delta E_Q$ of +2.76 mm/s and an asymmetry parameter $\eta$ of 0.6. Also for SIIb the calculated $\eta$ takes a rather large value. Assuming fast dynamic displacement between SIIa and SIIb, fast with respect to the lifetime of the exited state of the $^{57}$Fe nucleus (141 ns), we have to average the EFG tensors of both conformations before diagonalization. With equal populations of both conformations the averaged EFG tensor is

$$V(\text{SIIa}+\text{SIIb}) = [V(\text{SIIa})+V(\text{SIIb})] / 2 = \begin{vmatrix} +0.3 & +0.3 & -0.1 \\ +0.3 & +1.2 & 0.0 \\ -0.1 & 0.0 & -1.5 \end{vmatrix} \text{ a. u.,}$$

with the eigenvalues –1.53 a.u., +0.21 a.u., and +1.30 a.u. The resulting values for quadrupole splitting and asymmetry parameter are +2.64 mm/s and 0.7, respectively.

A further step of possible dynamic structural disorder includes the view that there are two orientations of the NO group in SII possible that differ by a 90° rotation around the axis perpendicular to the equatorial plane, as observed by X-ray studies [10] and by neutron diffraction studies [23]. Fast dynamic displacement from SIIa, via SIIb, to SIIa' (with the NO group being rotated by 90° with respect to SIIa) yields an averaged EFG tensor,

$$V(\text{SIIa}+\text{SIIa'}) = \begin{vmatrix} +0.7 & 0.0 & -0.1 \\ 0.0 & +0.7 & -0.1 \\ -0.1 & -0.1 & -1.4 \end{vmatrix} \text{ a. u.},$$

with eigenvalues –1.41 a.u., +0.70 a.u., and +0.70 a.u. Obviously this EFG tensor is axially symmetric and has a vanishing asymmetry parameter. The calculated quadrupole splitting (+2.26 mm/s), asymmetry parameter ($\eta = 0$), the positive sign of the quadrupole splitting, and the orientation of the EFG (along the molecular symmetry axis) are in reasonable agreement with experimental results. Another more complicated model for the dynamic behavior of the NO group suggests not only the previously discussed rotation but also a linearization of the bonds Fe−O−N by the displacement from SIIa, to SIIb. This is equivalent to an oscillation between SIIa and SIIb via the SI configuration. In summary, displacement conditions could be complicated as e.g. SIIa→SI→SIIb→SI→SIIa'→SI→…. . Beyond this so-called libration-rotation model even more complicated models including transitions via GS cannot be excluded.

### B. Temperature-dependent Mössbauer study of SII

The temperature-dependent Mössbauer study of SII performed in the range between 83 and 135 K (Fig. 2) is in favour of dynamic properties of SII. The normalized area of SII is reduced by about a factor of three when going from 83 to 135 K. This is not the result of a simultaneous decay process of SII to GS (which would occur at a temperature of about 150 K) because the normalized area of GS does not change significantly. The small decrease of the GS normalized area (inset of Fig. 2) of less than 15 % is in agreement with the temperature dependence of the Lamb-Mössbauer factor $f_{LM}$ in this temperature range.

The Mössbauer line area depends on the so-called effective thickness $t=n_A\sigma_0 f_{LM}$. The number of the $^{57}$Fe nuclei $n_A$ in the resonance absorber, the distribution of the NP ions between GS and SII (if decay process of SII into the GS does not take place), and the maximum value for resonance absorption cross-section $\sigma_0$ remain constant within the above-mentioned temperature-dependent study. The significant decrease of the normalized area of SII with increasing temperature therefore is subject to changes of $f_{LM}$. This behaviour corresponds to the proposed dynamic models for SII (*vide supra*). The motion of the NO group involves also the iron nucleus and, thus, increases its mean-square displacement $<x^2>$ and lowers the value of $f_{LM}$ of SII when increasing temperature.

### C. Nuclear inelastic scattering study of SII

The NIS spectra of NP anions in GS (open circles) and partly populated by SII (closed circles) are shown in Fig. 3a. In Fig. 3b the transfer of SI into SII and its depopulation into GS, stimulated by illumination with IR light at wave length 1064 nm is depicted (adapted from [16]). With a light exposure of 200 J/cm$^2$ a SII population of about 20-25 % is expected. This value is not reached in the present NIS experiment. The decrease of line intensity of the Fe–N stretching mode of GS at 83.6 meV and also of the Fe–C$_{eq}$ stretching mode of GS at 63.6 meV after illumination is only about 7 % (Fig. 3a), thus indicating that the population of SII can not be higher than this value. From DFT simulations a NIS peak at 67 meV is predicted for SII. If, however, this would be distributed over substates SIIa, SIIb and SIIts, it were broadened and further reduced in line intensity. The inevitable goal of future NIS experiments on SII, therefore, is to reach higher SII populations in SNP single crystals.

Fig. 3b suggests that optimal transfer of SI into SII is provided by a light exposure of 200 J/cm$^2$. This however, applies only for TMS but not for NIS measurements. The reason is (i) that TMS makes use of 14.4 keV Mössbauer transition and NIS of the characteristic X-ray, K$_\alpha$ radiation of 6.4 keV and (ii) that the mass absorption coefficient of SNP for 14.4 keV is lower than that for 6.4 keV, i.e. 15.1 and 33.2 cm$^2$/g [32]. Therefore the escape depth of the 6.4 keV radiation in an NIS experiment is restricted to a thin surface layer of about 100 μm thickness. It seems that this thin surface layer has been overexposed to light during the transfer from SI to SII, i.e. it first has reached optimal population of SII and then was depopulated (Fig. 3b) down to only 7 %. From a separate measurement of the SNP single crystal in nuclear forward scattering (NFS) geometry and detecting 14.4 keV radiation a volume population of SII of about 18 % was estimated. Obviously our experimental condition has caused high volume and low surface population of SII. Hence, we suggest that for future NIS measurements of SII in SNP the SI-to-SII transfer in the surface layer should be prepared at a much lower light exposure of about 20 J/cm$^2$.

## IV. CONCLUSION

On the basis of DFT calculations we suggest that the NO group in the SII state of NP is dynamically displaced over eclipsed and staggered orientations. More complicated dynamic behaviour discussed in terms of the libration-rotation model for the NO group can not be excluded. Mössbauer measurements performed in the temperature range 83 - 135 K are also in favour of dynamic properties of SII: the decrease of the Lamb-Mössbauer factor $f_{LM}$ for SII with increasing temperature is significant and leads to a strong reduction of the effective thickness and therefore of the normalized area of the quadrupole doublet of SII.

NIS studies were carried out at low surface population (~ 100 μm) of SII of about 7 %. The chosen light exposure of 200 J/cm$^2$ was optimal for volume population of SII and therefore for investigations by TMS. For NIS the transfer of SI to SII by the Nd-YAG laser should be performed with a much lower light exposure of about 20 J/cm$^2$. Temperature-dependent NIS measurements would elucidate further details of the dynamic structural properties of SII.

## ACKNOWLEDGMENTS

Financial support by the Deutsche Forschungsgemeinschaft (DFG) is gratefully acknowledged. We thank A. I. Chumakov and U. Ponkratz from the ESRF in Grenoble, France for assistance within the NIS investigation of SII.

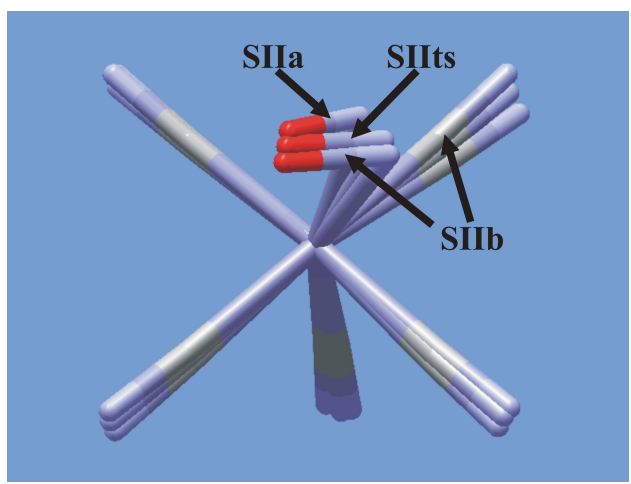

Fig. 1. Superposition of the geometries of SIIa, SIIb and SIIts (calculated with B3LYP/TZVP) constituting a dynamic model for SII. Note that in this dynamic model the equatorial CN and *trans*-CN groups perform bending vibrations out-of-phase simultaneously with the dynamic behaviour of the NO group.

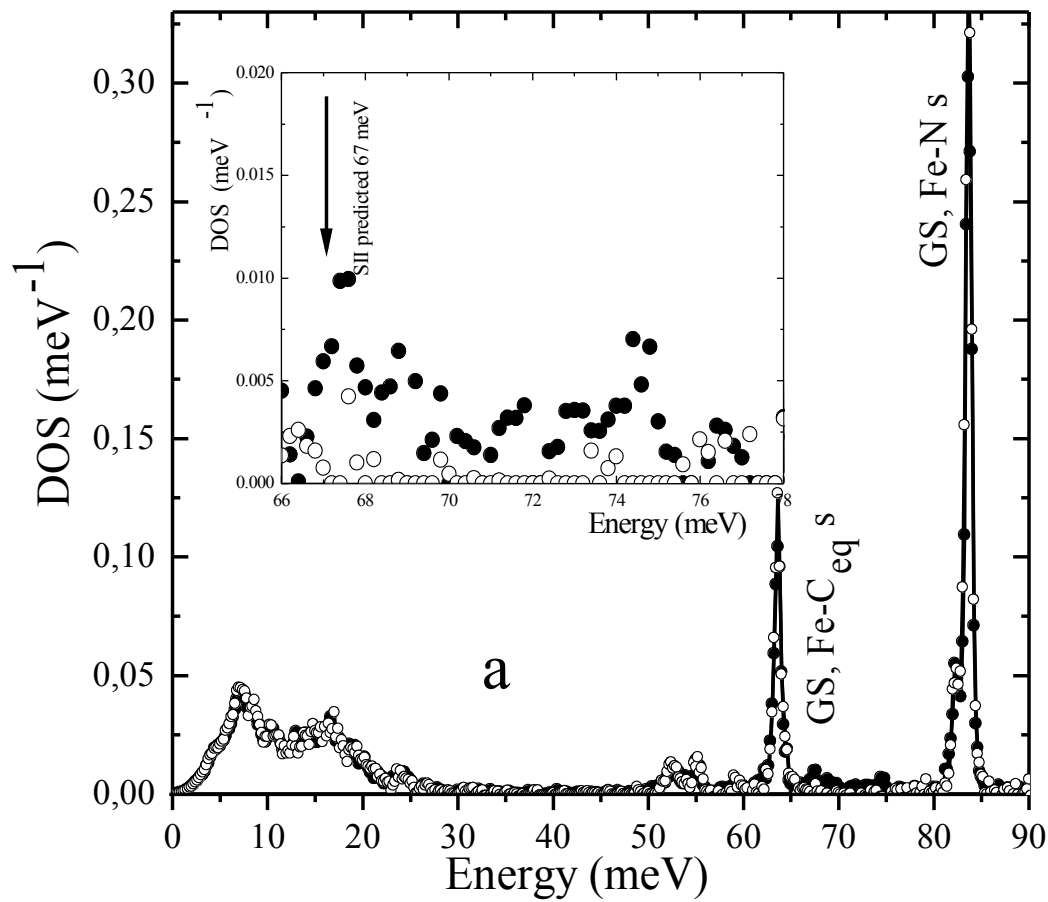
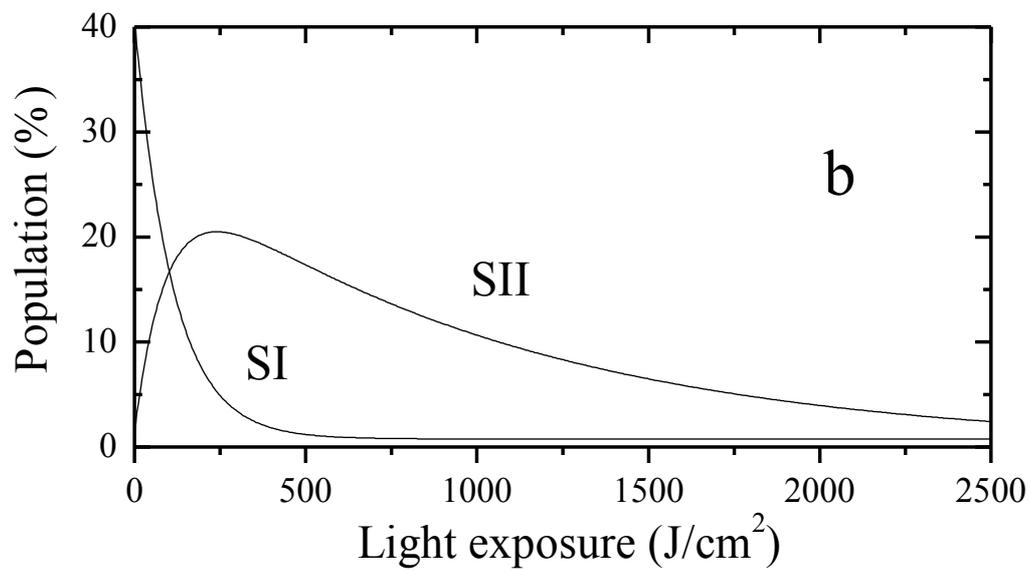

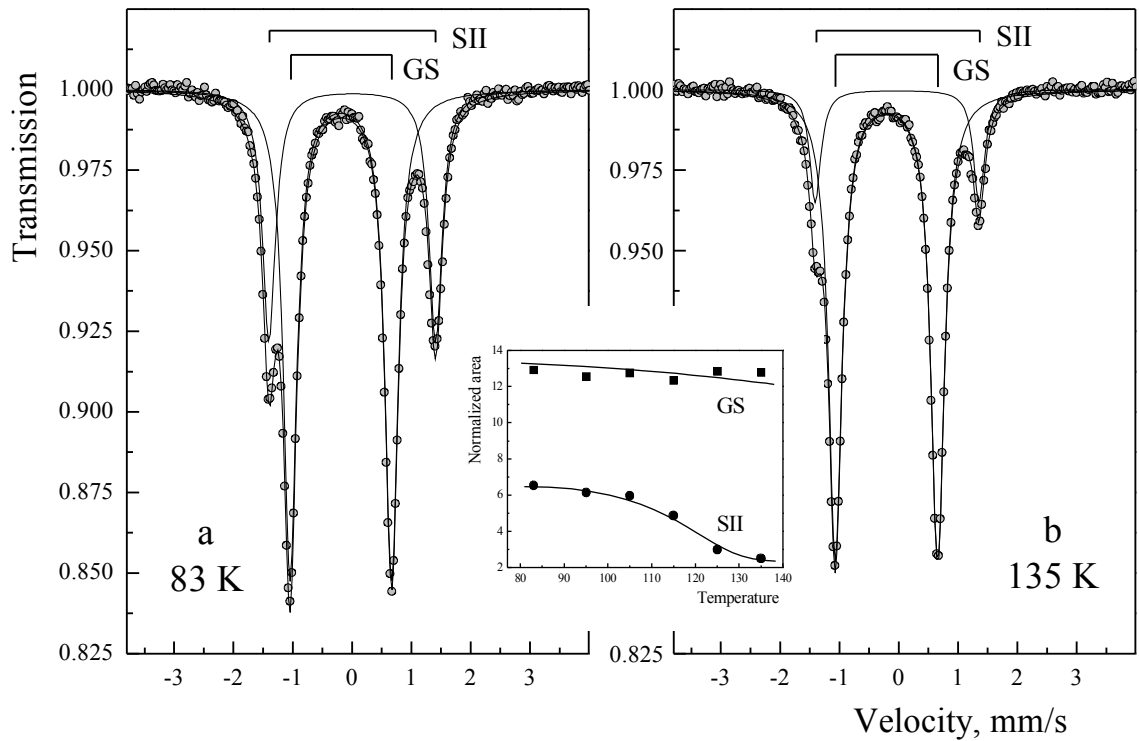

Fig. 2. Mössbauer spectra obtained at 83 K and 135 K from a SNP single crystal, **b**-cut, crystal thickness 490 μm and natural isotope abundance of iron, populated up to about 25 % with SII. In the inset the normalized areas for GS and SII are given.

Fig. 3. (a) Density of states (DOS) obtained by NIS at 98 K from a SNP single crystal enriched to 95 % with $^{57}$Fe. The synchrotron beam propagation direction is parallel to the **b** principal crystal axis. Solid circles: the crystal is to about 7 % populated with SII; open circles: after warming up to about 250 K for 15 minutes and cooling again to 98 K the metastable state SII is completely bleached and all NP anions are in the GS. The temperature was determined according to the so-called "detailed balance law": The excitation probability densities for phonon creation $S(E)$ and for annihilation $S(-E)$ are related by the Boltzmann factor, i.e. $S(-E) = S(E) \exp(-E/kT)$. The inset shows the enlarged part of the DOS between 66 and 78 meV.

(b) Transfer of SI to SII and depopulation to GS by illumination depending on light exposure (J/cm$^2$) with Nd-YAG laser, $\lambda$ = 1064 nm, **c**||**E**, **b**-cut. Taken from [16].